\documentclass[preprint,11pt]{elsarticle}
\usepackage{amsmath}
\usepackage{amssymb}
\usepackage{amsthm}
\usepackage{graphicx}
\usepackage{caption}
\usepackage{pdfpages}
\usepackage{adjustbox}
\usepackage{rotating}
\usepackage[T1]{fontenc}
\usepackage{amssymb}
\usepackage{mathrsfs}
\usepackage{multirow}
\usepackage{hhline}
\usepackage{enumerate}
\usepackage[algo2e]{algorithm2e}
\usepackage{algpseudocode}
\usepackage{etoolbox}
\usepackage{enumerate}
\usepackage{enumitem}
\usepackage{cite}
\usepackage{makecell}
\usepackage{algorithm}

\journal{}

\begin{document}
\renewcommand{\thesection}{\Roman{section}}
\newtheorem{theoremm}{Theorem}
\newtheorem{axiom}{Axiom}
\newtheorem{example}{Example}
\newtheorem{remarks}{Remarks}
\newtheorem {lemmaa}{Lemma}
\newtheorem{assumption}{Assumption}
\newtheorem {observation}[theoremm]{Observation}
\newtheorem {defnn}{Definition}
\newtheorem{proposition}{Proposition}
\newdefinition{remark}{Remark}
\newtheorem {corollary}{Corollary}
\newtheorem {conjecturee}[theoremm]{Conjecture}
\newtheorem {fact}[theoremm]{Fact}
\newtheorem {procd}{Procedure}
\newtheorem {rules}{Rule}
\begin{frontmatter}

\title{Temporally Non-Uniform Cellular Automata: Surjectivity, Reversibility and Cyclic Behavior}

\author[]{Subrata Paul\corref{cor1}}
\ead{subratapaul.sp.sp@gmail.com}

\author[]{Sukanta Das}
\ead{sukanta@it.iiests.ac.in}

\cortext[cor1]{Corresponding author}
\address{Department of Information Technology, Indian Institute of Engineering Science and Technology, Shibpur, India}

\begin{abstract}

This work studies Temporally Non-Uniform Cellular Automata (t-NUCAs), a variant of non-uniform cellular automata, which temporally use two rules in a sequence during their evolution. The one-dimensional t-NUCAs, under finite as well as infinite lattices, are considered in this work. Surjectivity and injectivity of the t-NUCAs are studied. The reversibility of the t-NUCAs is also explored here. Finally, a study on the cyclic behavior of finite t-NUCAs is presented.

\end{abstract}

\begin{keyword}
Temporally Non-Uniform Cellular Automata (t-NUCAs)\sep Surjectivity \sep Injectivity \sep Reversibility  \sep Recurrent configuration \sep Cyclic behavior

\end{keyword}

\end{frontmatter}

\section{Introduction}

A Cellular Automaton (CA) is a discrete dynamical system that consists of a regular network of cells. The cells update their states depending on the current states of the cell itself and the neighboring cells. Classically, a CA uses a single local rule to update the states of all cells synchronously during its evolution~\citep{BhattacharjeeNR16}. Recently, a class of non-classical (popularly called {\em non-uniform}) cellular automata have been introduced in~\citep{paul2024temporally}, in which different rules at different time steps are used to update their cells. This work formally defines this class of automata and studies their behavior.

Non-uniformity in cellular automata was first introduced in 1986~\citep{Pries86}, which allows the cells to use different rule. These non-uniform CAs were characterized by a number of researchers with a target to use them in VLSI design \& test and other related application domains, see for example \citep{BhattacharjeeNR16,ppc1,tcad/DasS10}. Later, this class of non-uniform CAs have been generalized and defined in~\citep{CattaneoDFP09,dennunzio2012non}. Dynamical behavior of these CAs, such as injectivity, surjectivity, permutivity, equicontinuity, decidability, structural stability, etc. have been investigated \citep{CattaneoDFP09,dennunzio2012non,dennunzio2014three}. Another kind of non-uniformity, called asynchronism, in cellular automata has also been studied, which do not force all the cells to be updated simultaneously \citep{naka}. A number of variations of such CAs have been proposed over the time and their dynamical behavior have been studied \citep{Fates14,Alberto12,Manzoni2012,Formenti2013,Sarkar12,fates2018}.

However, in all the above cases, the local rules remain invariant during evolution of the automaton. The present work deals with a different class of non-uniformity, in which the local rules can vary temporarily during the automaton's evolution. This class of automata is named temporally non-uniform cellular automata (t-NUCAs). In literature, we find that such a CA was first used to solve the density classification problem \citep{Fuk05}. Recently, some dynamical behavior of a special class of t-NUCAs has been studied \citep{paul2024temporally,ghosh2023}. A sub-class of t-NUCAs, named {\em temporally stochastic cellular automata}, in which two rules are stochastically used for evolution, have also been studied \citep{roy2022atsca,paul2023pattern}. This work develops a general framework of these CAs and formally defines the t-NUCAs. Like all of those works, the present work also concentrates on the one dimensional cases and considers that only two rules, say $f$ and $g$, are applied in a sequence. Such a t-NUCA is represented as $(f,g)[\mathcal{R}]$, where $\mathcal{R}$ is the sequence of rules used by the t-NUCA.

The presentation of a rule sequence, which is preferably an infinite sequence, is an issue. The issue is addressed in Section~\ref{Sec:RuleSeq}. Indeed, the rule sequences play a pivotal role in the dynamical behavior of t-NUCAs. So, the rule sequences are studied and utilized throughout the work. In the study of surjectivity of a t-NUCA $(f,g)[\mathcal{R}]$, the surjectivity of the CAs with rules $f$ and $g$ play a decisive role (see Section~\ref{Sec:SurInj}). The same is true in the study of injectivity. In the entire study, lattice of both types -- finite and infinite are considered. If the lattice type is not explicitly mentioned in any discussion, then the discussion is applicable to both the types.

To observe the reversible behavior in a t-NUCA, bijectivity of the t-NUCA is not a sufficient condition. The rule sequence is also to be {\em reversible}. Section~\ref{Sec:Reversibility} discusses the reversibility issue of a t-NUCA. It shows that the inverse of a t-NUCA is also a t-NUCA. Like any finite CA, a finite t-NUCA forms cycles in its configuration space. Section~\ref{rev_finite} throws some light on the cyclic behavior of the finite t-NUCAs. It characterizes a class of configurations, called {\em recurrent configurations}, which form cycles in the configuration space of a finite t-NUCA. Finally, Section~\ref{Sec:Conclusion} concludes the paper with pointers to the open questions and possible extension of the work.

\section{The Model}
\label{model}

A cellular automaton (CA) consists of a collection of cells that are arranged on a $D$-dimensional lattice ($D\ge 1$). A cell assumes a state from a finite set of possible states $S=\{0,1,\cdots, d-1\}$ ($d\ge 2$) that is updated in every time step following a {\em local transition function} (also called a {\em local rule}). All the cells of a CA use a single rule ($f$) for the update. The local rule $f$ takes the present states of the cell's neighbors to generate the next state. In classical cellular automata, the rule $f$ remains invariant temporarily during the evolution. This work deals with a class of non-classical CAs, where this constraint is compromised.

\subsection{Definition}

A {\em temporally Non-Uniform Cellular Automaton} (t-NUCA) is allowed to apply different local rules on the cells at different time steps. So, a rule sequence $(\mathcal{R}_t)_{t\geq 0}$, instead of a single rule, is required to define the CA, where $\mathcal{R}_t$ is the local rule applied at time step $t$. Formally, a t-NUCA can be defined as a quadruple $(\mathcal{L}, \mathcal{N}, \mathcal{S}, \mathcal{R})$, where 
\begin{enumerate}
	\item $\mathcal{L}\subseteq\mathbb{Z}^D$ is a $D$-dimensional lattice. Elements of $\mathcal{L}$ are the cells of the t-NUCA.
	\item $\mathcal{N}=(\vec{v}_1,\vec{v}_2,\cdots,\vec{v}_m)$ is the neighborhood vector of each cell $\vec{v}$ of the lattice, such that $(\vec{v}+\vec{v}_i)\in\mathcal{L}$.
	
	\item  $\mathcal{S}$ is the finite set of states that the cells can assume.
	
	\item  $\mathcal{R}=(\mathcal{R}_t)_{t\geq 0}$ is the sequence of local rules, temporarily applied on the cells. That is, a rule $\mathcal{R}_t :\mathcal{S}^m\rightarrow \mathcal{S}$ of the sequence is applied to all the cells at time $t$. If $s^t_{\vec{v}}\in \mathcal{S}$ is the state of the cell $\vec{v}\in \mathcal{L}$ at $t$, then the next state of the cell is $s^{t+1}_{\vec{v}}=\mathcal{R}_t(s^t_{\vec{v} + \vec{v}_1},s^t_{\vec{v} + \vec{v}_2},\cdots,s^t_{\vec{v} + \vec{v}_m})$. 
\end{enumerate}

In principle, a t-NUCA can be defined on any dimension $D$ and can use any set of local rules during its evolution. However, the present work considers the following.
\begin{itemize}
\item $D=1$. That is, the t-NUCAs under consideration are defined over one-dimensional lattice. The lattice can be finite as well as infinite. A t-NUCA is called finite if the lattice is finite. For finite t-NUCAs, periodic boundary condition is assumed here.
\item Only two local rules, say $f$ and $g$, are used in the rule sequence $\mathcal{R}$ of a t-NUCA. Such a t-NUCA is represented as $(f,g)[\mathcal{R}]$. 
\end{itemize}

A configuration of a t-NUCA is an assignment of states to all the cells. It is represented as $c=(s_{\vec{v}})_{\vec{v}\in\mathcal{L}} \in\mathcal{S}^\mathcal{L}$. We denote $\mathcal{C}=\mathcal{S}^{\mathcal{L}}$ as the set of all configurations of the CA. During evolution, a t-NUCA hops from one configuration to another. The local rule $f$ of a t-NUCA $(f,g)[\mathcal{R}]$ induces a map $G_f:\mathcal{C}\rightarrow\mathcal{C}$, such that
$$G_f(\vec{v})\:|_{\vec{v}\in\mathcal{L}}=f(s_{\vec{v} + \vec{v}_1},s_{\vec{v} + \vec{v}_2},\cdots,s_{\vec{v} + \vec{v}_m})$$
Similarly, another map $G_g$, induced by the rule $g$, can be obtained. These two maps are applied temporarily on $\mathcal{C}$ to get the {\em global transition function} ($G$) of the t-NUCA. 
That is, the $G:\mathcal{C}\rightarrow\mathcal{C}$ is obtained from $G_f$ and $G_g$, where for any configuration $c\in\mathcal{C}$,

\begin{align}
\label{Eqn:Global}
	G(c)&=\begin{cases}
		G_f(c) & \text{if the rule $f$ is applied}\\
		G_g(c) & \text{if the rule $g$ is applied}
	\end{cases}	
\end{align}

\noindent A t-NUCA can also be identified by its global transition function $G$. Similarly, a classical CA that uses only rule $f$ ($g$) for its evolution is identified by $G_f$ ($G_g$). In subsequent discussions, the $G_f$ (resp. $G_g$) implies the global transition function of the classical CA with rule $f$ (resp. $g$). 

The following example illustrates the evolution of a t-NUCA. In this example, as well as in most of the subsequent examples, elementary cellular automata (ECAs) rules are taken. An ECA is one-dimensional, binary, 3-neighborhood CA. As per Wolfram's convention, an ECA rule $f$ is represented by a `decimal code'  $w = f(0,0,0)\cdot2^0 + f(0,0,1)\cdot 2^1 + \cdots + f(1,1,1)\cdot 2^7$. 

\begin{figure}[ht]
	\begin{center}		
		\begin{tabular}{ccc}
			\includegraphics[width=100 mm]{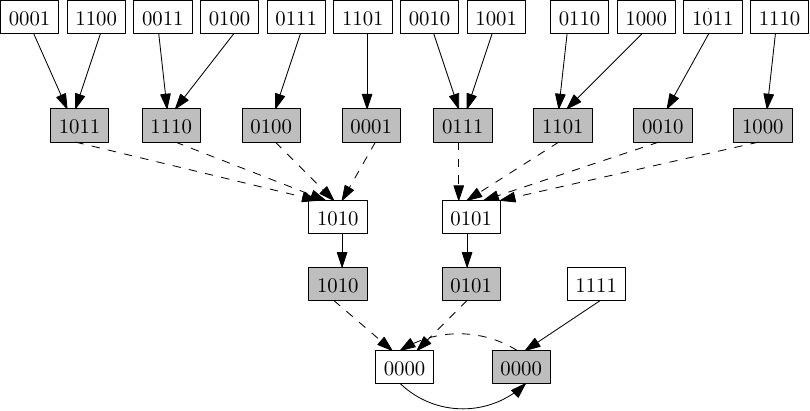}& \hspace{1 em}&				
		\end{tabular}		
		\caption{Transition diagram of the t-NUCA of Example~\ref{Example:tNUCA3090}} 	
		\label{fig:30_90_10}
	\end{center}
\end{figure}

\begin{example}	
	\label{Example:tNUCA3090}	
	Let us consider a one-dimensional finite t-NUCA $(30,90)[\mathcal{R}]$ with 4-cells that uses periodic boundary condition. That is, the t-NUCA uses the ECA rules 30 and 90 for evolution. Suppose, the rule sequence is $\mathcal{R}=(\mathcal{R}_t)_{t\ge 0}$, where $\mathcal{R}_t=30$ if $t$ is even, $\mathcal{R}_t=90$ otherwise. Therefore, the rule sequence is $30,90,30,90,\cdots$. Now, if the initial configuration of the t-NUCA is 0001, it moves to 1011 in next step by applying rule 30. Next, the CA reaches 1010 using rule 90, and so on. Finally, the t-NUCA settles at 0000. The transitions are shown in Figure~\ref{fig:30_90_10}, where solid and dashed arrows represent the application of rule 30 and rule 90 respectively. Figure~\ref{fig:30_90_10} notes down transitions for all possible initial configurations, which are shown in white boxes. The configurations, marked by gray boxes, are the intermediate configurations that are reached during evolution from initial configurations. Observe that some configurations are repeated in the transition diagram with different trajectories. 
\end{example} 

\subsection{Rule sequence}
\label{Sec:RuleSeq}
The presentation of a rule sequence $\mathcal{R}=(\mathcal{R}_t)_{t\ge 0}$ of a t-NUCA $(f,g)[\mathcal{R}]$ is an issue. If a well-defined condition, say $\Theta(t)$, is identified, which is true when the rule $f$ is applied, then the sequence can be expressed as following:
\begin{align}\label{eq:4}
	\mathcal{R}_t &=
	\begin{cases}
		f  &  \text{ if }\Theta(t) \text{ is true }\\
		g  & \text{ otherwise }\
	\end{cases}
\end{align}
The rule sequence can also be presented exploiting $\Theta(t)$ only. Following are some of such presentations.
\begin{itemize}
	\item {\em Integer sequence representations:} If we note down the time $t$ when $\Theta(t)$ is true, then a sequence of integers is obtained. These integers indicate the time steps when $f$ is applied. For example, if $\Theta(t)$ is true when $t$ is odd, then we get the following series: $1,
	3,5,7,9,\cdots$. The t-NUCA can be represented as: $(f,g)[1,3,5,7,9,\cdots]$.
	
	\item {\em OEIS number representation:} If a series of integers, as above, is found to represent a rule sequence, then the rule sequence can alternatively be presented by an {\em OEIS number}. In the database of OEIS~\citep{oeis}, the integer sequences are designated by some lexicographical ordering number. For example, the above series ($1,3,5,7,9,\cdots$) is described by OEIS number $A005408$. Then, the t-NUCA can be presented as $(f,g)[A005408]$.
	
	\item {\em Binary sequence representation:} A direct representation of a rule sequence is the use of a binary sequence, in which 1 and 0 indicate the application of the rule $f$ and the rule $g$ respectively. Hence, the above rule sequence (with integer sequence: $1,3,5,7,9,\cdots$) can be presented as $010101\cdots$. Therefore, the t-NUCA can be represented as $(f,g)[0101010\cdots]$. Since, in the sequence $010101\cdots$, the term `$01$' repeats over and over, the sequence can be presented as  $(01)^+$. Then, the t-NUCA is denoted as $(f,g)[(01)^+]$.	
\end{itemize}

\begin{figure}[ht]
	\begin{center}
		\begin{tabular}{ccc}
			ECA $90$ & ECA $73$ & $(90,73)$[$(110)^+$] \\[6pt]
			\includegraphics[width=31mm]{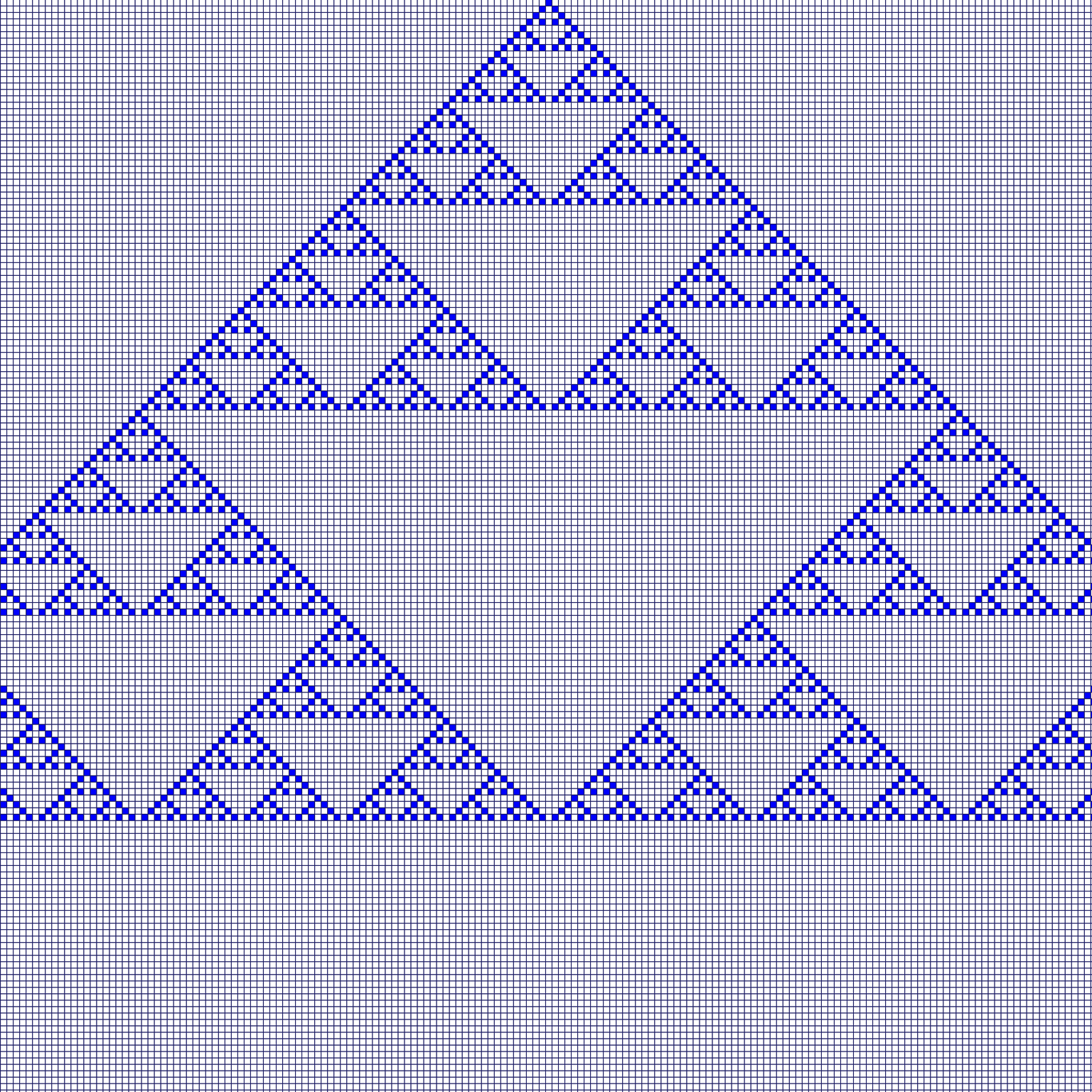} & \includegraphics[width=31mm]{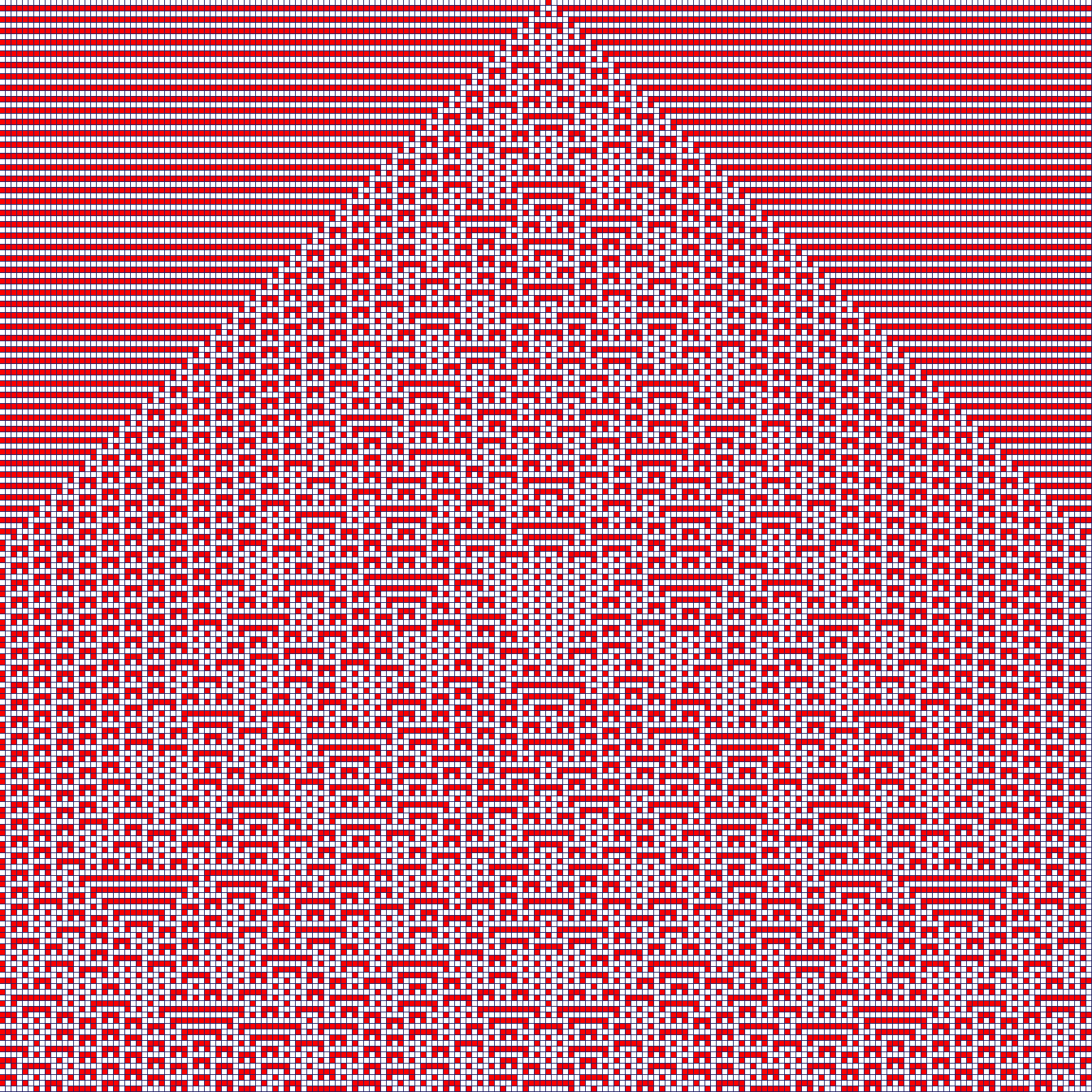} & \includegraphics[width=31mm]{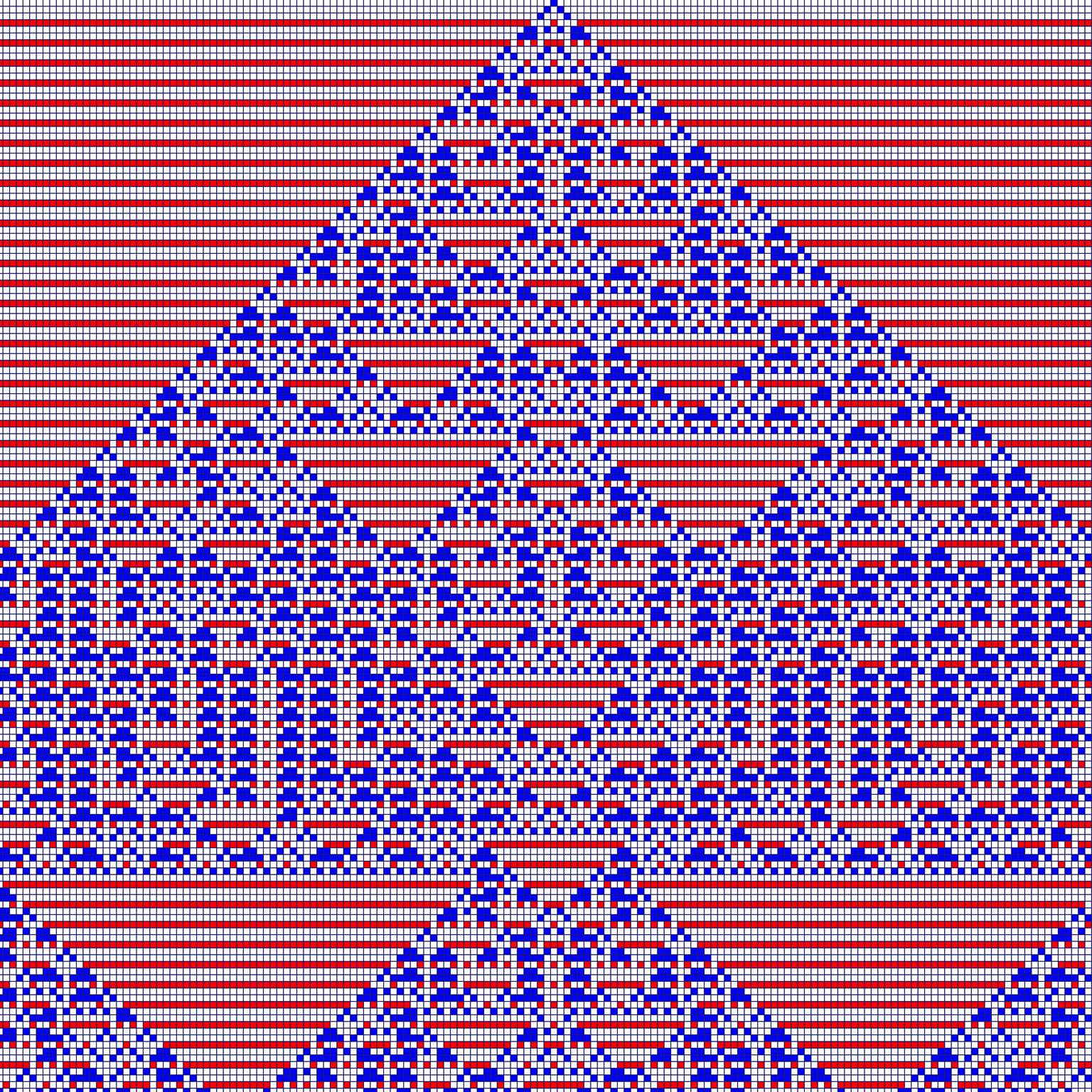} \\
    	\end{tabular}
		\caption{Space-time diagram of ECA $90$, ECA $73$ and t-NUCA $(90,73)[(110)^+]$}
		\label{Fig:Type1}
	\end{center}
\end{figure}

\noindent In the present work, the binary sequence presentation is mostly used. The rule sequences, however, can be classified as {\em periodic} and {\em aperiodic}. A rule sequence is periodic if a finite subsequence of rules repeats infinitely in the sequence. The t-NUCA $(30,90)[101010\cdots]$ of Example~\ref{Example:tNUCA3090} uses a periodic rule sequence. As another example, consider the t-NUCA $(90,73)[110110110\cdots]$. It also uses a periodic rule sequence, in which the subsequence 110 (that is, the rule subsequence (90, 90, 73)) repeats infinitely. The space-time diagram of the t-NUCA is shown in Figure~\ref{Fig:Type1}. Here, the cells with state 1 that are updated by $f$ ($g$) are marked by {\em blue} ({\em red}), whereas the cells with state $0$ are marked by {\em white}. The difference in dynamical behavior of the t-NUCA from ECA 90 and ECA 73 can also be observed in Figure~\ref{Fig:Type1}.

A rule sequence, which is not periodic, is {\em aperiodic}. For example, consider the t-NUCA $(90,73)[\mathcal{R}]$ with $\mathcal{R}=(\mathcal{R}_t)_{t\ge 0}$ where 
\begin{align}\label{eq:7}
	\mathcal{R}_t &=
	\begin{cases}
		90 & \text{if $t$ is not prime}\\
		73  & \text{otherwise}\\
	\end{cases}
\end{align}

\begin{figure}[ht]
	\begin{center}
		\begin{tabular}{cc}
			$(90,73)$[$A018252$] & $(164,131)$[$A018252$]  \\[6pt]
			\includegraphics[width=40mm]{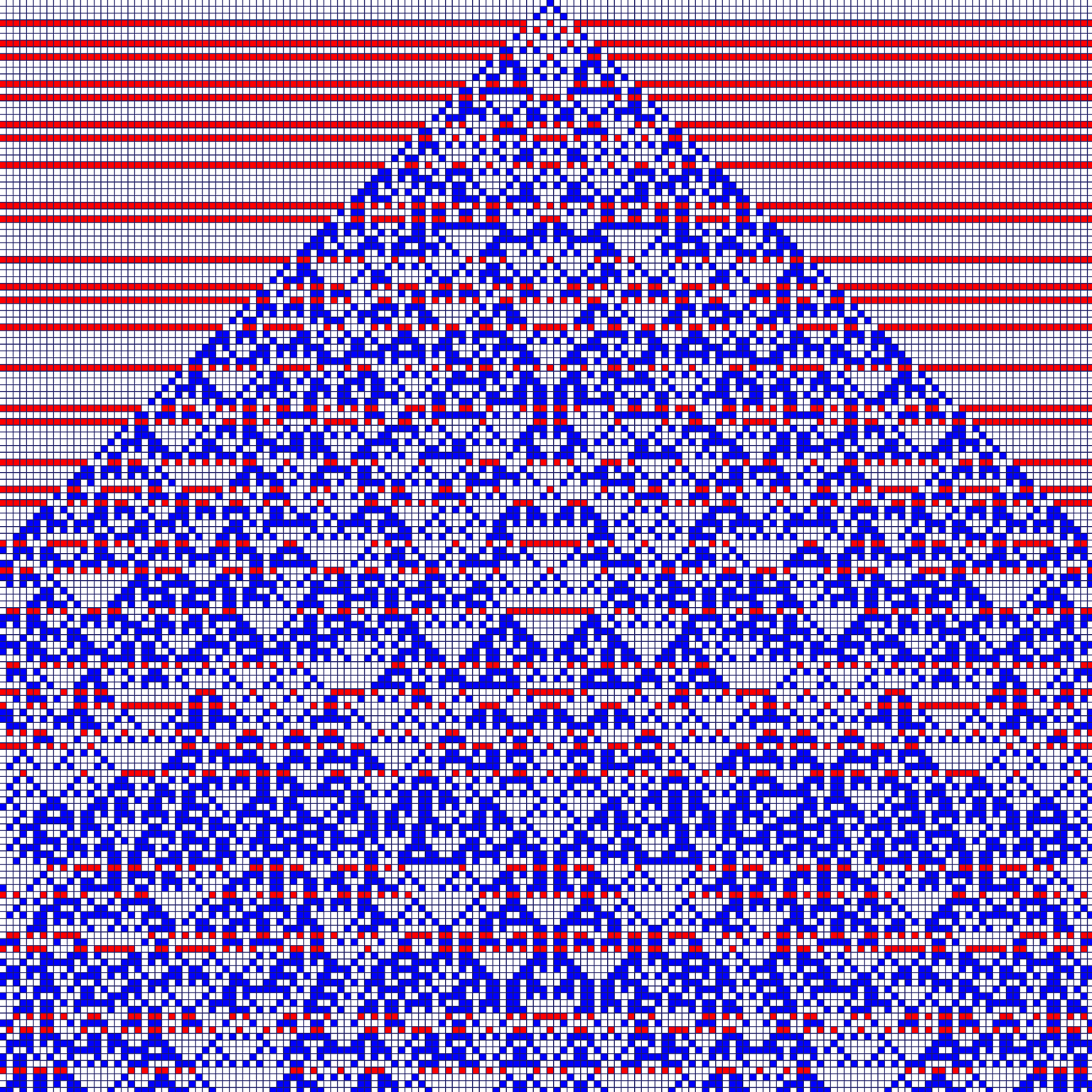} & \includegraphics[width=40mm]{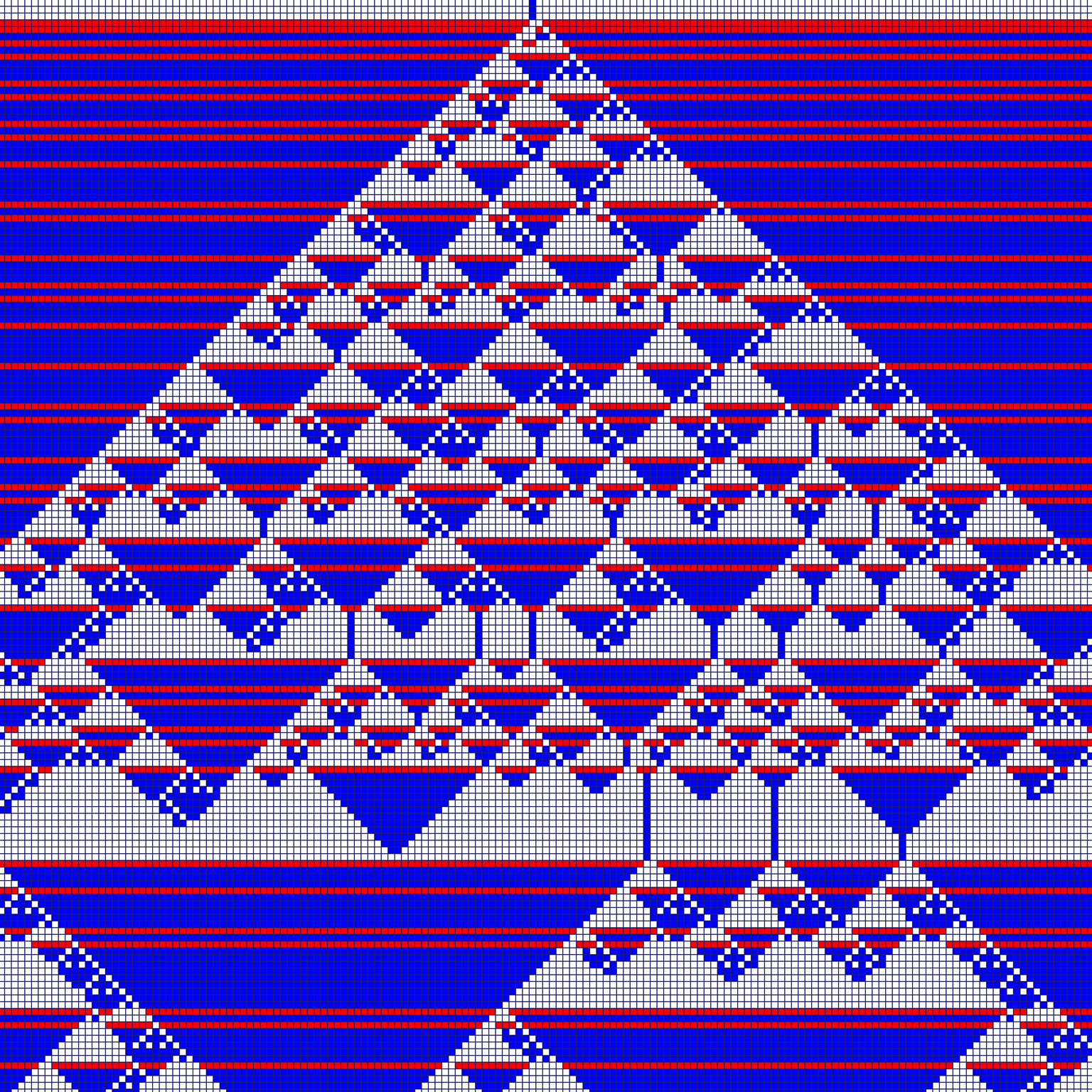} \\ 			
		\end{tabular}
		\caption{Space-time diagrams of the t-NUCAs $(90,73)[A018252]$ and $(164,131)[A018252]$}
		\label{Fig:Type2}
	\end{center}
\end{figure}

\noindent Hence, $\Theta(t)$ is true for the following $t$ -- $0,1,4,6,8,9,10,\cdots$ (OEIS: $A018252$). This is an aperiodic sequence. Figure~\ref{Fig:Type2} shows the space-time diagrams of the t-NUCAs $(90,73)[A018252]$ and $(164,131)[A018252]$. 

The condition $\Theta(t)$ can be a deterministic function, as well as a stochastic function. In case of stochastic function, the rule sequence is generally different in different runs, and the sequence is aperiodic.

\section{Surjectivity and Injectivity}
\label{Sec:SurInj}

A classical CA with rule $f$ is surjective if its global transition function $G_f$ is surjective. A surjective CA has no {\em non-reachable} (Garden-of-Eden) configurations. In other word, the non-existence of non-reachable configurations implies that the CA is surjective. We use this property to define surjectivity of a t-NUCA and investigate the existence of {\em non-reachable} configurations in a t-NUCA through the $G_f$ and $G_g$.

\begin{defnn}
	A configuration of a t-NUCA is non-reachable if the t-NUCA can not reach the configuration during its evolution from any initial configuration. A t-NUCA is surjective if it has no non-reachable configurations.
\end{defnn}

The configurations 1111, 0011, 0110, 1100 and 1001 of the t-NUCA of Figure~\ref{fig:30_90_10} are {\em non-reachable}. Hence, the t-NUCA is non-surjective.

\begin{theoremm}\label{Theorem:Non-reachableCommon}
	There exists a non-reachable configuration in a t-NUCA $(f,g)[\mathcal{R}]$ if $X_f\cap X_g\neq \emptyset$, where $X_f$ and $X_g$ are the sets of non-reachable configurations of $G_f$ and $G_g$, respectively.
\end{theoremm}

\begin{proof}
Let us consider a t-NUCA $(f,g)[\mathcal{R}]$ where $X_f$ and $X_g$ are the sets of non-reachable configurations of $G_f$ and $G_g$ respectively, and $X_f \cap X_g \neq \emptyset$. Now, if $c\in X_f\cap X_g$ then $c$ can not be reached from any configuration by either rule in $\mathcal{R}$. Hence, $c$ is a non-reachable configuration of the t-NUCA.
\end{proof}

In ECAs 30 and 90 with 4 cells, the 1100 is a common non-reachable configuration. So, the configuration 1100 is non-reachable in a 4-cell t-NUCA involving these two rules (see Figure~\ref{fig:30_90_10}). However, sometimes it is possible to identify a configuration as non-reachable by observing the presence of an {\em orphan}.

\begin{defnn}
A finite subconfiguration is called {\em orphan} if it does not appear in the next configuration of any configuration of a CA, defined over an infinite lattice.
\end{defnn}

In other words, an orphan has no preimage. That is, the presence of an orphan in a configuration makes it non-reachable. In classical CAs under an infinite lattice, a non-reachable configuration is identified through an orphan \citep{Kari2009,Amoroso72}.

\begin{corollary}
\label{Corollary:NonSurjectiveInfinite}
Let $G_f$ and $G_g$ both have atleast one orphan.
\begin{itemize}
\item[I:] Let the lattice $\mathcal{L}=\mathbb{Z}$. Then, a t-NUCA $(f,g)[\mathcal{R}]$ is non-surjective for any rule sequence $\mathcal{R}$.
\item[II:] Let the lattice $\mathcal{L}$ be finite. Then, there exists a positive integer $n_0$ for which a finite t-NUCA $(f,g)[\mathcal{R}]$ with $n\ge n_0$ cells is non-surjective for any rule sequence $\mathcal{R}$.
\end{itemize}
\end{corollary}

\begin{proof}
Let $x$ and $y$ be two orphans of $G_f$ and $G_g$ respectively.\\
{\em Case I:} Hence, $G_f$ and $G_g$ both are non-surjective when the lattice $\mathcal{L}=\mathbb{Z}$. Since $x$ and $y$ are finite, a configuration $c\in\mathcal{C}$ can be found which contains both of $x$ and $y$. Hence, the $c$ cannot be reached by $G_f$ or by $G_g$ due to the presence of $x$ and $y$ respectively. Hence, $c$ is a non-reachable configuration of a t-NUCA $(f,g)[\mathcal{R}]$ for any rule sequence $\mathcal{R}$. That is, the t-NUCA $(f,g)[\mathcal{R}]$ is non-surjective.\\

\noindent {\em Case II:}  Let $x$ and $y$ be the orphans of minimum length among the other orphans of $G_f$ and $G_g$ respectively. Since $x$ and $y$ are finite, there exists an array of size $n_0$, which accommodates $x$ and $y$ optimally. Now, in a finite t-NUCA of $n\ge n_0$ cells, a configuration $c$ can be found which contains $x$ and $y$ both. Hence, the configuration $c$ cannot be reached neither by $G_f$ nor by $G_g$. Therefore, the $c$ is non-reachable in the t-NUCA. That is, the t-NUCA is non-surjective. Hence the proof.
\end{proof}

\begin{figure}[ht]
\centering
\includegraphics[width=75mm]{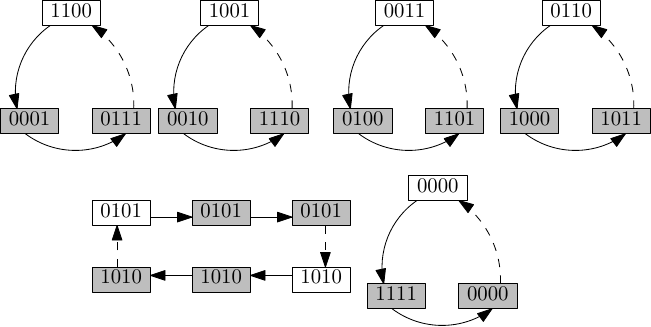}
\caption{Transitions in the $4$-cell t-NUCA $(7,40)[(110)^+]$}
\label{fig:7X40}

\end{figure}

Let us consider two non-surjective CAs -- ECA 7 and ECA 40 under infinite lattice. Let us use the Amaroso-Patt algorithm \citep{Amoroso72} to find orphans of these two CAs. The orphans are 01001 and 0111, respectively. In the t-NUCA $(7,40)[\mathcal{R}]$, with any rule sequence $\mathcal{R}$, there exist non-reachable configurations, under infinite lattice, that contain both the orphans as subconfigurations. Hence, the t-NUCA is non-surjective. When the lattice is finite, with at least 9 cells, the t-NUCA contains non-reachable configurations that contain the orphans as subconfigurations. Therefore, the t-NUCA with $n\ge 9$ cells is non-surjective. However, a t-NUCA with $n<n_0$ cells may behave different. For example, the t-NUCA $(7,40)[(110)^+]$ with 4 cells is surjective. See Figure~\ref{fig:7X40} for verification.

However, Theorem~\ref{Theorem:Non-reachableCommon} does not state the sufficient condition for existence of non-reachable configuration in a t-NUCA. There exists non-reachable configuration even if $G_f$ and $G_g$ do not have any non-reachable configurations in common (that is, $X_f\cap X_g=\emptyset$). Following example illustrates the case.

\begin{figure}
	\begin{tabular}{c}		
		\includegraphics[scale=0.7]{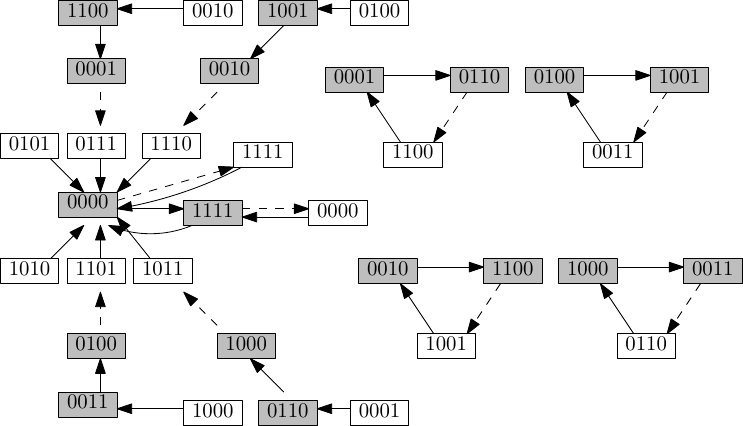}
	\end{tabular}
	\centering
	\caption{Transition diagram of a 4-cell t-NUCA $(3,15)[(110)^+]$}
	\label{fig:3X15}	
\end{figure}

\begin{example}\label{ex5}
	Let us consider a 4-cell t-NUCA $(3,15)[(110)^+]$, where 3 and 15 are two ECA rules (boundary condition is periodic). Figure~\ref{fig:3X15} shows the transition diagram of the t-NUCA with solid and dashed arrows indicating the transitions by rule 3 and rule 15 respectively. The configurations in the diagram, marked by gray boxes, are the intermediate configurations that are reached during evolution from initial configurations (marked by white boxes). Since the 4-cell ECA 15 is surjective, the ECAs do not have non-reachable configurations in common; that is, $X_{3} \cap X_{15}=\emptyset$. But, the t-NUCA has non-reachable configurations.
\end{example}

\begin{theoremm}\label{Theorem:SurjectiveOne}
There exists a surjective t-NUCA $(f,g)[\mathcal{R}]$ for a rule sequence $\mathcal{R}$ if $G_f$ or $G_g$ is surjective. 
\end{theoremm}

\begin{proof}
If  $G_f$ is surjective and the rule $f$ is used first in the sequence, then all configurations of the t-NUCA are reachable. Hence, a rule sequence, putting $f$ as the first rule, can be found for which the t-NUCA $(f,g)[\mathcal{R}]$ is surjective. Similarly, if $G_g$ is surjective, then another rule sequence with $g$ as the first rule of the sequence can be found which makes the t-NUCA surjective. Hence the proof.
\end{proof}

\begin{figure}[h]
\centering
\includegraphics[scale=0.6]{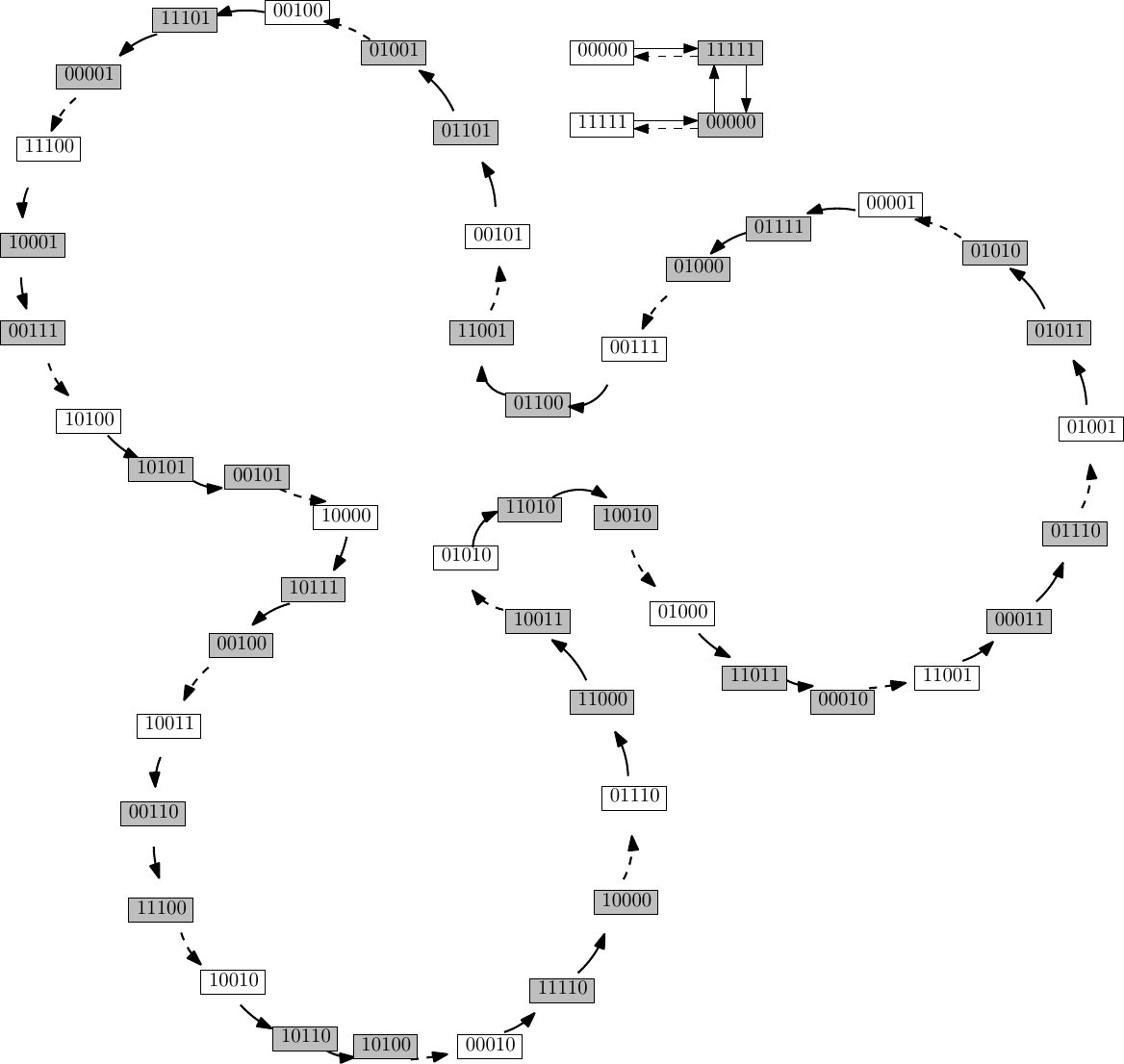}
\caption{The transitions of the 5-cell t-NUCA $(15,25)[(110)^+]$}
\label{fig:15X25}
\end{figure}

\begin{example}
Let us consider a 5-cell t-NUCA $(15,25)[(110)^+]$, where rule 15 is used first in the sequence. Here, the 5-cell ECA 15 is surjective but the 5-cell ECA 25 is not. However, the t-NUCA is surjective. Figure~\ref{fig:15X25} illustrates the fact by depicting the transitions of the t-NUCA. The solid (resp. dashed) arrows of the figure represent the transitions by rule 15 (resp. rule 25). Observe that the t-NUCA has no non-reachable configuration. When defined over infinite array, the t-NUCA $(15,25)[(110)^+]$ remains surjective. Because, ECA 15 is surjective and ECA 25 is non-surjective while defined over infinite lattice.
\end{example}

\begin{corollary}
\label{Corollary:1stRuleSurjective}
A t-NUCA $(f,g)[\mathcal{R}]$ is surjective if $f$ (resp. $g$) is the first rule in the rule sequence and $G_f$ (resp. $G_g$) is surjective.
\end{corollary}
\begin{proof}
The proof directly follows from the argument of the proof of Theorem~\ref{Theorem:SurjectiveOne}.
\end{proof}

\begin{corollary}
\label{Corollary:SurjectiveBoth}
For any rule sequence $\mathcal{R}$, the t-NUCA $(f,g)[\mathcal{R}]$ is surjective if $G_f$ and $G_g$ are surjective.
\end{corollary}

\begin{proof}
Since each configuration is reachable by the $G_f$ and $G_g$ both, there does not exist non-reachable configuration for any rule sequence $\mathcal{R}$. 
\end{proof}

\begin{figure}
	\centering
	\includegraphics[scale=0.6]{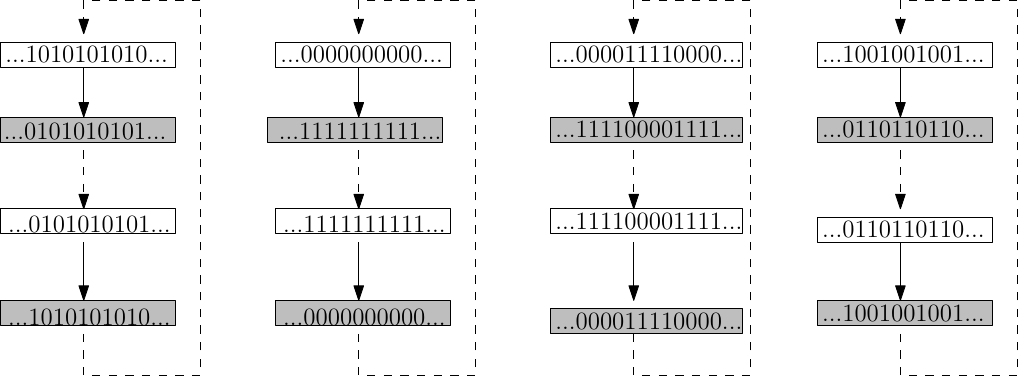}
	\caption{Some transitions of t-NUCA ($51$, $204$)[$(01)^+$]}
\label{fig:inf}
\end{figure}

\begin{example}	
\label{Example:Infinite51204}
	Let us consider two surjective CAs - ECA 51 and ECA 204, defined over infinite lattice. Consider a t-NUCA $(51,204)[(01)^+]$ with these two rules, which is also defined on infinite lattice. The t-NUCA is surjective. Figure~\ref{fig:inf} illustrates the fact by showing some transitions of the t-NUCA with the solid and dashed arrows indicating the transitions by rule 51 and rule 204 respectively. A t-NUCA with these two rules and another rule sequence (such as $(110)^+$) is also surjective.
\end{example}

Let us now summarize the above discussion on surjectivity. Following two statements are true for any type of lattice (finite and infinite).
\begin{enumerate}
\item If $G_f$ and $G_g$ are surjective, then a t-NUCA $(f,g)[\mathcal{R}]$ is surjective for any rule sequence $\mathcal{R}$.
\item If $G_f$ is surjective and $G_g$ is non-surjective, then a t-NUCA $(f,g)[\mathcal{R}]$ is surjective for any rule sequence $\mathcal{R}$ that contains rule $f$ as the first rule.
\end{enumerate}
However, if $G_f$ and $G_g$ have orphans, then for any rule sequence $\mathcal{R}$, a t-NUCA $(f,g)[\mathcal{R}]$ is non-surjective (1) under infinite lattice, and (2) under finite lattice of size $n\ge n_0$, for some $n_0\ge 1$.

Hence, the case of deciding surjectivity of a t-NUCA when the $G_f$ is non-surjective but $G_g$ is surjective, and $f$ is applied first, is left. In addition to this, another case is left: surjectivity of a t-NUCA $(f,g)[\mathcal{R}]$ when  $G_f$ and $G_g$ are surjective under infinite lattice, but non-surjective under finite lattice of size $n$. We need to develop decision algorithms, if any exist, to deal with the above cases.

To decide surjectivity of one-dimensional classical CAs (under infinite lattice), Amaroso and Patt proposed an algorithm in their seminal paper \citep{Amoroso72}. For a periodic rule sequence $\mathcal{R}$, the same algorithm can be extended to decide surjectivity of a t-NUCA $(f,g)[\mathcal{R}]$ under infinite lattice when $G_f$ is non-surjective but $G_g$ is surjective, and rule $f$ is applied first. Following are the steps of the decision algorithm.
\begin{itemize}
\item[{\em Step 1:}] Let the period length of $\mathcal{R}$ be $p$ and $\mathcal{R}=(\mathcal{R}_0,\mathcal{R}_1,\cdots,\mathcal{R}_{p-1})^+$. Set $h_0\leftarrow \mathcal{R}_0$ and $i\leftarrow 0$.
\item[{\em Step 2:}] Test the surjectivity of the CA with rule $h_i$ using the Amaroso-Patt algorithm. If CA is surjective, then report the t-NUCA $(f,g)[\mathcal{R}]$ as surjective, and return.
\item[{\em Step 3:}] $i\leftarrow i+1$. If $i=p$, then report the t-NUCA as non-surjective. Otherwise, construct a rule $h_i$ which is equivalent to the consecutive application of $h_{i-1}$ and $\mathcal{R}_i$, and go to {\em Step 2}. 
\end{itemize}

\noindent In the above construction of $h_i$, if the radius of $h_{i-1}$ and $\mathcal{R}_i$ are $r'$ and $r$ respectively, then the radius of $h_i$ is $r'+r$. The radius of a rule defines the neighbors of a cell on each side. Example~\ref{Example:SurjectiveAlgo} shows such construction. However, the correctness of the algorithm follows from the argument of the proof of Corollary~\ref{Corollary:NonSurjectiveInfinite}: If each $h_i$ is non-surjective, then there exist an orphan, say $x_i$, in the CA. Now, a configuration $c$ can be found which includes each of the $x_0,x_1,\cdots, x_{p-1}$. Hence, $c$ is a non-reachable configuration to each of the CA with rule $h_i$, $i\in\{0,1,\cdots,p-1\}$. Since the rule sequence is periodic and such $h_i$ can be obtained, so the t-NUCA is non-surjective. On the other hand, if any one of the new CAs with rule $h_i$ is surjective, then obviously the t-NUCA is surjective.

\begin{theoremm}
The surjectivity of one-dimensional t-NUCAs with periodic rule sequences is decidable.
\end{theoremm}

\begin{proof}
For finite lattice, the problem is always decidable. For infinite lattice, the decidability directly follows from the above algorithm.
\end{proof}

\begin{example}
\label{Example:SurjectiveAlgo}
Let us test surjectivity of a t-NUCA $(22,105)[(10)^+]$ over an infinite lattice where $G_{22}$ is non-surjective and $G_{105}$ is surjective. As per the above algorithm, a rule $h_1$, equivalent to the consecutive application of the ECA rules 22 and 105, is to be found out. For the shake of clarity, both the ECA rules are shown in Table~\ref{Table:look-up}. If a cell gets its state after application of rules 22 and 105, a 5-bit string in a configuration dictates the state of the cell. As example, consider the string of states in cells $i-2,i-1,i,i+1,i+2$ is 00101. After applying rule 22 on the configuration, the cells $i-1,i,i+1$ get the states 110 . And, the state of cell $i$ in next step, after applying rule 105, is 1. Hence, $h_1(00101)=1$. In this way, the rule $h_1$ can be identified (see Table~\ref{Table:look-up-h}). The construction of $h_1$ proves that the transformation of a configuration by consecutive application of rules 22 and 105 is equivalent to the transformation of the same configuration by the rule $h_1$.
	\begin{table}
		\centering
		\caption{The ECA rules 22 and 105}
		\scalebox{0.9}{
			\begin{tabular}{|l|l|l|l|l|l|l|l|l|l|}
				\hline
				$xyz\in\{0,1\}^3$ & 111 & 110 & 101 & 100 & 011 & 010 & 001 & 000 & $f$ \\ \hline
				$f(xyz)$   & 0   & 0   & 0   & 1   & 0   & 1   & 1   & 0   & 22   \\ \hline
				$f(xyz)$  & 0   & 1   & 1   & 0   & 1   & 0   & 0   & 1   & 105  \\ \hline
		\end{tabular}}
	\label{Table:look-up}
	\end{table}
	
	\begin{table}
		\centering
		\caption{The rule $h_1$, constructed from rules 22 and 105}
		\scalebox{0.52}{
			\begin{tabular}{|l|l|l|l|l|l|l|l|l|l|l|l|l|l|l|l|l|}
				\hline
				$uvxyz$ & 01111 & 01110 & 01101 & 01100 & 01011 & 01010 & 01001 & 01000 & 00111 & 00110 & 00101 & 00100 & 00011 & 00010 & 00001 & 00000 \\ \hline
				$h_1$ & 1     & 1     & 1     & 0     & 0     & 1     & 0     & 1     & 0     & 0     & 1     & 0     & 0     & 1     & 0     & 1     \\ \hline
				$uvxyz$ & 11111 & 11110 & 11101 & 11100 & 11011 & 11010 & 11001 & 11000 & 10111 & 10110 & 10101 & 10100 & 10011 & 10010 & 10001 & 10000 \\ \hline
				$h_1$ & 1     & 1     & 1     & 0     & 1     & 0     & 1     & 0     & 1     & 1     & 0     & 1     & 1     & 0     & 1     & 0     \\ \hline
			\end{tabular}
		}
		
		\label{Table:look-up-h}
	\end{table}

Now, the surjectivity of $h_1$ is to be tested by Amaroso-Patt algorithm, which shows that the $h_1$ is surjective. Hence, as per the algorithm, the t-NUCA is surjective. To verify the result, let us identify the orphans of ECA 22. The orphans are 010101001 and 010010101. Hence, any configuration $c$ that contains any of these strings is non-reachable. However, these two strings have preimage in $h_1$. Hence, $c$ is reachable by $h_1$. That is, the t-NUCA is surjective.
\end{example} 

Let us now extend the definition of injectivity of classical CAs to the t-NUCAs. In a t-NUCA, a configuration can be reached by $G_f$ as well as by $G_g$. Hence, to define injectivity of a t-NUCA, which rule (either $f$ or $g$) is used to get the successor of a configuration is to be considered. That is, the injectivity of a t-NUCA is defined through the injectivity of $G_f$ and $G_g$.

\begin{defnn}
\label{Definition:Injectivity}
A t-NUCA $(f,g)[\mathcal{R}]$ is injective iff $G_f$ and $G_g$ are injective.
\end{defnn}

The t-NUCA of Example~\ref{Example:tNUCA3090} is not injective, as the CA with rule 30 ($G_{30}$) is not injective. Observe that $G_{30}(0001)=G_{30}(1100)$ (Figure~\ref{fig:30_90_10}). On the other hand, the t-NUCA of Example~\ref{Example:Infinite51204} is injective, as both the CAs with rules 51 and 204 are injective. Each configuration of this t-NUCA has unique predecessor when rule 51 (or 204) is applied (Figure~\ref{fig:inf}).

%

Hence, the decidability about injectivity of t-NUCAs depends on that of $G_f$ and $G_g$. Since injectivity of one-dimensional CAs over infinite lattice is decidable \citep{Amoroso72}, injectivity of one-dimensional t-NUCA is also decidable. 

A t-NUCA is bijective iff the t-NUCA is injective and surjective. A t-NUCA $(f,g)[\mathcal{R}]$ is injective iff $G_f$ and $G_g$ both are injective. And, in classical CAs, injectivity implies surjectivity \citep{KARI20053}. Further, when $G_f$ and $G_g$ both are surjective, the t-NUCA is also surjective  (Corollary~\ref{Corollary:SurjectiveBoth}). That is, if a t-NUCA is injective, the t-NUCA is surjective. Hence, a t-NUCA is bijective iff the t-NUCA is injective.\\ 


\section{Reversibility}
\label{Sec:Reversibility}
Reversibility is a property of some mathematical and physical systems in which time-reversed dynamics is well-defined. That is, a reversible system can move in backward direction with respect to time. A t-NUCA can also show time-reversal dynamics. A t-NUCA is called reversible if the evolution of the t-NUCA can be inverted with respect to time.

\begin{defnn}
A t-NUCA with global transition function $G$ is reversible if there exists a function $H$ such that, for any configuration $x\in\mathcal{C}$, when $y=G(x)$ then $x=H(y)$. Here, $H$ is the inverse of $G$ --that is, $H=G^{-1}$.
\end{defnn}

If a t-NUCA is injective, then each configuration, with respect to the applied rule of the t-NUCA, has a unique predecessor. So, if the {\em reverse} of the rule sequence ($\mathcal{R}$) of an injective t-NUCA $(f,g)[\mathcal{R}]$ is known, then the time-reversal behavior of the t-NUCA can be observed. That is, unlike classical CAs, the injectivity is not sufficient to decide reversibility of a t-NUCA. If the {\em reverse} of $\mathcal{R}$ does not exist, the time-reversal dynamics cannot be observed in the injective t-NUCA. 

If a sequence is $abc$, then its reverse sequence is $cba$. The reverse of a finite sequence can be obtained by reading it from the opposite side. However, the rule sequence of a t-NUCA is preferably infinite. An infinite sequence cannot be read from the opposite side. So, for serving our purpose, we define an infinite rule sequence as reversible in the following way.

\begin{defnn}
A rule sequence $\mathcal{R}=(\mathcal{R}_t)_{t\ge 0}$ is called reversible if there exists another rule sequence $\mathcal{K}=(\mathcal{K}_t)_{t\ge 0}$, such that for any $q\in\mathbb{N}$, either ($\mathcal{K}_0\mathcal{K}_1\cdots\mathcal{K}_q$) is the reverse of ($\mathcal{R}_0\mathcal{R}_1\cdots\mathcal{R}_q$), or there exists an $\epsilon>0$ for which ($\mathcal{K}_0\mathcal{K}_1\cdots\mathcal{K}_{q+\epsilon}$) is the reverse of ($\mathcal{R}_0\mathcal{R}_1\cdots\mathcal{R}_{q+\epsilon}$).
\end{defnn}

For example, the rule sequence $(110)^+$ in binary is reversible. The $(011)^+$ is the reverse sequence. If $q=2$, 011 is the reverse of 110. If $q=3$, 0110 is not the reverse of 1101, but 011011 is the reverse of 110110 ($\epsilon =2$).

The $\Theta$ of Equation~\ref{eq:4} can be considered as a generating function which generates the rule sequence $\mathcal{R}=(\mathcal{R}_t)_{t\ge 0}$ in binary. In case of aperiodic rule sequences, if $\Theta$ serves as a probability function, then the rule sequence cannot be reversed. Hence, the t-NUCA with such aperiodic rule sequence is {\em irreversible}.
The periodic rule sequences, on the other hand, are reversible. Let the period of a periodic rule sequence be $p$. As per Equation~\ref{eq:4}, the sequence can be expressed (in binary) as $(\Theta(0)\Theta(1)\cdots\Theta(p-1))^+$ where $\Theta(t)=1$ if $\Theta(t)$ is true and $\Theta(t)=0$ otherwise. Then, its reverse sequence is $(\Theta(p-1)\Theta(p-2)\cdots\Theta(0))^+$.

\begin{theoremm}
A t-NUCA $(f,g)[\mathcal{R}]$ is reversible iff the t-NUCA is injective and $\mathcal{R}$ is a reversible sequence.
\end{theoremm}

\begin{proof}
If a t-NUCA $(f,g)[\mathcal{R}]$ is injective, then $G_f$ and $G_g$ are both injective. Since injectivity implies bijectivity in classical CAs, $G_f$ and $G_g$ are bijective. Now, for any configuration $x\in\mathcal{C}$, if $y=G(x)$ and the $\mathcal{R}$ can be reversed, then with the reversed rule sequence and using $G_f^{-1}$ and $G_g^{-1}$, $x$ can be obtained from $y$. That is, there exists a function that acts as the inverse of $G$ such that $x=G^{-1}(y)$. Hence, the t-NUCA is reversible. On the contrary, if the t-NUCA is reversible, then the t-NUCA is to be injective and reversed rule sequence of the $\mathcal{R}$ has to exist. Hence the proof.
\end{proof}

The t-NUCA $(51,204)[(01)^+]$ of Example~\ref{Example:Infinite51204} is reversible. The $G_{51}$ and $G_{204}$ both are injective, hence the t-NUCA is injective. Further, the reverse of the rule sequence $(01)^+$ exists, which is $(10)^+$. Hence, the t-NUCA is reversible. This can also be verified from Figure~\ref{fig:inf}.

\begin{theoremm}
The inverse of a reversible t-NUCA is a t-NUCA.
\end{theoremm}

\begin{proof}
Let $(f,g)[\mathcal{R}]$ be a reversible t-NUCA. Hence, a reversed rule sequence of $\mathcal{R}$ exist. Suppose, the reversed rule sequence is $\mathcal{K}=(\mathcal{K}_t)_{t\ge 0}$. If $G$ is the global transition function of the t-NUCA, then the inverse of $G$ can be found as following:
\begin{align}\label{eq:12}
G^{-1}(c)&=\begin{cases}
G^{-1}_f(c) & \text{if $\mathcal{K}_t=f$}\\
G^{-1}_g(c) & \text{otherwise}
\end{cases}
\end{align}
Since $G_f$ and $G_g$ are injective (hence bijective), $G^{-1}_f$ and $G^{-1}_g$ exist. By Hedlund theorem \citep{hedlund69}, $G^{-1}_f$ and $G^{-1}_g$ are also CAs. Suppose, local rules of these two CAs are $f'$ and $g'$ respectively. These two rules are involved in the t-NUCA $(f',g')[\mathcal{K}]$. If $G'$ is the global transition function of this t-NUCA, then from Equation~\ref{eq:12}, we get that
\begin{align}
G'(c)=G^{-1}(c)&=\begin{cases}
G_{f'}(c)=G^{-1}_f(c) & \text{if $\mathcal{K}_t=f'$}\\
G_{g'}(c)=G^{-1}_g(c) & \text{otherwise}
\end{cases}
\end{align}
That is, the t-NUCA $(f',g')[\mathcal{K}]$ shows the time-reversed evolution of the t-NUCA $(f,g)[\mathcal{R}]$. Hence the proof.
\end{proof}

\begin{figure}[ht]
\begin{center}		
\begin{tabular}{c}
	\includegraphics[scale=0.8]{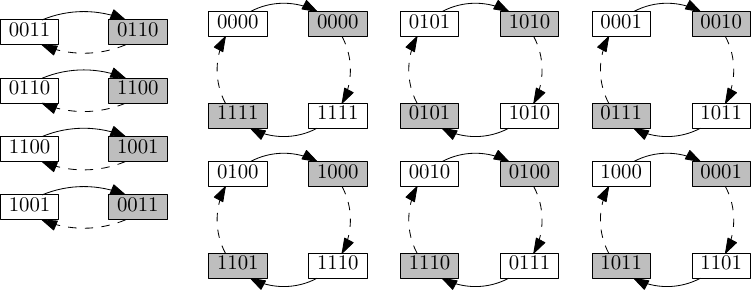}
\end{tabular}

\caption{Transition diagram of 4-cell reversible t-NUCA ($170$, $85$)[$(10)^+$]}
\label{fig:reversible}
\end{center}
\end{figure}

\begin{example}	\label{e6}
Let us consider a 4-cell t-NUCA $(170,85)[(10)^+]$ with ECA rules 170 and 85. The ECAs 170 and 85 with 4 cells are injective, and the reverse of the rule sequence $(10)^+$ is $(01)^+$. Hence, the t-NUCA is reversible. Figure~\ref{fig:reversible} shows the transition diagram of the t-NUCA. The inverse CAs of ECAs 170 and 85 are ECAs 240 and 15 respectively. Hence, the inverse t-NUCA of the t-NUCA $(170,85)[(10)^+]$ is $(240,15)[(01)^+]$. The transition diagram of the t-NUCA $(240,15)[(01)^+]$ is identical to Figure~\ref{fig:reversible} with inverted arrows.
\end{example}

Deciding reversibility of a t-NUCA depends on two factors -- the reversibility of $G_f$ and $G_g$, and the reversibility of the rule sequence. In one-dimension, reversibility of $G_f$ and $G_g$ is decidable. So, reversibility of one-dimensional t-NUCA is decidable if the rule sequence is reversible. However, the reversibility of two or higher dimensional t-NUCAs is undecidable as the reversibility of $G_f$ and $G_g$ is undecidable when the dimension of the CAs is two or more \citep{Kari90}. 

\section{Cycles in finite t-NUCAs}\label{rev_finite}

Let us now turn our attention to the finite lattice only. A finite t-NUCA forms cycles in its configuration space. A closer look to the transition diagrams of finite t-NUCAs reveals that some configurations repeat infinitely in cycles. Let us call these configurations as {\em recurrent}, by taking vocabulary from {\em Markov Chain} theory. Similarly, there are configurations which do not recur infinitely during evolution of a t-NUCA. They are {\em transient} configurations. In a transition diagram of a finite t-NUCA, a configuration may appear more than once. So, there is a possibility that one configuration recurs infinitely for an initial configuration, but for another initial configuration, the configuration appears only finite number of times. We treat this configuration as recurrent.

\begin{defnn}
A configuration $x\in \mathcal{C}$ of a t-NUCA is {\em transient} if it is either non-reachable or, when $x$ is reachable for an initial configuration, then it is visited only a finite number of times and does not recur afterwards. In contrast, a configuration is {\em recurrent} if there exists an initial configuration for which $x$ is not transient but recurs infinitely many times during evolution of the t-NUCA.
\end{defnn}

In Figure~\ref{fig:3X15}, configuration 0101 is transient, whereas 1100 is recurrent. Observe that for the initial configuration 0010, the configuration 1100 does not recur infinitely. But, it recurs infinitely for the initial configuration 1001, which makes the 1100 recurrent. 

To facilitate our next study, let us introduce the left shift operator ($\sigma$) on the rule sequence. Let $\mathcal{R}=(\mathcal{R}_t)_{t\ge 0}$ be a rule sequence and $\mathcal{R}'$ be its $m$-place left shift. Then, the new rule sequence is the following.
\begin{equation}
\mathcal{R}'=\sigma^m(\mathcal{R})=\sigma^m((\mathcal{R}_t)_{t\ge 0})=(\mathcal{R}_t)_{t\ge m}
\end{equation}

For example, if $1010010001\cdots$ is a rule sequence in binary, then the rule sequence with its 3-place left shift is $\sigma^3(1010010001\cdots)=001000100001\cdots$. However, if a rule sequence is left shifted, the recurrent configurations remain recurrent. 

\begin{theoremm}\label{Theorem:recurrentConfiguration}
A recurrent configuration of an $n$-cell t-NUCA $(f,g)[\mathcal{R}]$ is recurrent in the $n$-cell t-NUCA $(f,g)[\sigma^m(\mathcal{R})]$ for any $m\ge 0$.
\end{theoremm}

\begin{proof}
Let $(f,g)[\mathcal{R}]$ be an $n$-cell t-NUCA, generates a sequence of configurations $x_2,x_3,\cdots$ from an initial configuration $x_1$. Assume that $x_k$ ($k\ge 1$) is a recurrent configuration of the t-NUCA. Let us now consider a rule sequence $\mathcal{R}'=\sigma^m(\mathcal{R})$ for any $m\ge 0$. In $(f,g)[\mathcal{R}]$, the configuration sequence $x_1,x_2,\cdots,x_m,\cdots$ is generated due to the application of the rule sequence $\mathcal{R}_1,\mathcal{R}_2,\cdots,\mathcal{R}_m,\cdots$. Now, in the t-NUCA, if $\mathcal{R}$ is replaced by $\sigma^m(\mathcal{R})$, then the new t-NUCA (that is, $(f,g)[\sigma^m(\mathcal{R})]$) can generate the same sequence of configurations from $x_m$. As $x_k$ is recurrent and so appears infinitely many times in the sequence $x_1,x_2,\cdots$, it appears infinitely many times in the sequence $x_m,x_{m+1},\cdots$. That is, $x_k$ remains recurrent. Hence the proof.
\end{proof}

The recurrent configurations are visited many times during evolution, hence form cycles in the transition diagram. Let us call the time required to recur a configuration as {\em recurrent length}. Then, for a recurrent configuration, we can get a sequence of {\em recurrent lengths}. Let $\Lambda_x$ be the sequence of recurrent lengths of the t-NUCA that evolves from $x$. The $\Lambda_x$ is an infinite sequence when $x$ is a recurrent configuration. However, $\Lambda_x$ is either empty or a finite sequence if $x$ is transient. If $\Lambda_x$ is not empty for a transient configuration $x$, then a cycle is formed involving $x$. But, the cycle itself is transient as it does not appear after leaving $x$ permanently. Whereas, the cycles involving $x$ recur infinitely when $x$ is a recurrent configuration.

In the 4-cell t-NUCA $(170,85)[(10)^+]$ of Figure~\ref{fig:reversible}, the 1111 is a recurrent configuration with $\Lambda_{1111}=(1,3,1,3,1,3,\cdots)$. 
In some cases, a subsequence of recurrent lengths repeats in the sequence. In the above example, the lengths $(1,3)$ repeat in $\Lambda_{1111}$. Hence, this sequence is periodic. The following is an interesting result.

\begin{theoremm}
	The sequence of recurrent lengths in $\Lambda_x$ of a recurrent configuration $x$ is periodic if the rule sequence is periodic.
\end{theoremm}
\begin{proof}
	Let $\Lambda_x=(\lambda_1,\lambda_2,\cdots)$. Suppose the recurrent length $\lambda_1$ is obtained by applying a subsequence $s_1$ of rules, $\lambda_2$ by the next subsequence $s_2$ and so on. So, the rule sequence for this case is $S=(s_1,s_2,\cdots)$. Suppose, $S$ is periodic. Then $s_1$ is periodically observed. Since $x$ is a recurrent configuration, at some time the CA is to be at configuration $x$ before following the rule subsequence $s_1$. Then the recurrent length $\lambda_1$ can be observed. Subsequently, the recurrent lengths $\lambda_2,\lambda_3,\cdots$ are observed. This makes $\Lambda_x$ periodic.
	
\end{proof}

The recurrent configurations of a t-NUCA form cycles. Unlike the cycles of classical CAs, a cycle of t-NUCA may contain a configuration multiple times. In a t-NUCA with periodic rule sequence, $\Lambda_x=(\lambda_1,\lambda_2,\cdots,\lambda_l)^+$ for a recurrent configuration $x$, and hence $\lambda_1+\lambda_2+\cdots+\lambda_l$ ($=len$, say) is fixed. Such a fixed $len$ cannot be found for a recurrent configuration of a t-NUCA with aperiodic rule sequence.

\begin{defnn}
Two rule sequences $\mathcal{R}$ and $\mathcal{R}'$ are {\em shift equivalent} to each other if there exists $k_1,k_2 \ge 0$, such that $\mathcal{R}=\sigma^{k_1}(\mathcal{R}')$ and $\mathcal{R}'=\sigma^{k_2}(\mathcal{R})$. 
\end{defnn}

For example, the periodic rule sequences $(1101)^+$ and $(1011)^+$ are shift equivalent to each other as $(1101)^+=\sigma^3((1011)^+)$ and $(1011)^+=\sigma^1((1101)^+)$. In case of periodic rule sequences, two sequences are shift equivalent to each other if the period of one sequence is the circular left shifts of the period of the other sequence. Hence, for any periodic rule sequence, a shift equivalent rule sequence can be found which is also periodic. Following results can be derived from Theorem~\ref{Theorem:recurrentConfiguration}.

\begin{corollary}
\label{Corollary:RecurrentShiftEq}
A configuration of an $n$-cell t-NUCA $(f,g)[\mathcal{R}]$ is recurrent iff it is recurrent in the $n$-cell t-NUCA $(f,g)[\mathcal{R}']$, where $\mathcal{R}$ and $\mathcal{R}'$ are shift equivalent to each other.
\end{corollary}

\begin{proof}
Since $\mathcal{R}$ can be obtained by the left shifts of $\mathcal{R}'$ and vice versa, the corollary directly follows from Theorem~\ref{Theorem:recurrentConfiguration}.
\end{proof}

\begin{corollary}
A configuration of an $n$-cell t-NUCA $(f,g)[\mathcal{R}]$ is transient iff it is transient in the $n$-cell t-NUCA $(f,g)[\mathcal{R}']$, where $\mathcal{R}$ and $\mathcal{R}'$ are shift equivalent to each other.
\end{corollary}

\begin{proof}
By Corollary~\ref{Corollary:RecurrentShiftEq}, all the recurrent configurations of an $n$-cell t-NUCA $(f,g)[\mathcal{R}]$ are recurrent in the $n$-cell t-NUCA $(f,g)[\mathcal{R}']$ and vice versa, where $\mathcal{R}$ and $\mathcal{R}'$ are shift equivalent to each other. Hence, all the transient configurations of the t-NUCA $(f,g)[\mathcal{R}]$ remain transient in the t-NUCA $(f,g)[\mathcal{R}']$. The converse is also true. Hence the proof.
\end{proof}

\begin{theoremm}
All configurations of a finite reversible t-NUCA are recurrent.
\end{theoremm}

\begin{proof}
Let us consider an $n$-cell reversible t-NUCA $(f,g)[\mathcal{R}]$. By contradiction, let us assume that there exists a configuration, say $x_0$, which is not recurrent but transient. Now, if the t-NUCA runs from $x_0$, then a sequence of configurations, say $x_0,x_1,\cdots,x_i,\cdots$, is generated. Since the lattice is finite, the configurations of the sequence repeat after some time steps. Let $x_i$ repeats as $x_j$ ($i\ne j$). That is, $x_i=x_j$. As $x_0$ is assumed as transient, so $i>0$. Now, the $x_i$ has two predecessors -- $x_{i-1}$ and $x_{j-1}$. Hence, a many-to-one mapping which disallows the t-NUCA to go backward direction of time. It is a contradiction as the t-NUCA is reversible. Therefore, our assumption is wrong, which implies that the $x_0$ is recurrent. Hence the proof.
\end{proof}

The 4-cell t-NUCA $(170, 85)[(10)^+]$ of Example~\ref{e6} is reversible. All the configurations of the t-NUCA are recurrent (see Figure~\ref{fig:reversible}). However, an irreversible t-NUCA may contain only recurrent configurations, see Figure~\ref{fig:15X25} for example.

In case of finite t-NUCA with periodic rule sequence, the recurrent and transient configurations can be figured out. The recurrent lengths, hence possible cycle lengths, of a recurrent configuration can also be found out. However, for aperiodic rule sequences, such a decision may not be taken always.

\section{Conclusion}
\label{Sec:Conclusion}

The paper has formally defined the t-NUCAs and explored the set theoretic properties and cyclic behavior of one dimensional cases. In a t-NUCA $(f,g)[\mathcal{R}]$, the rule sequence $\mathcal{R}$ plays a pivotal role in its dynamical behavior. This work has studied the t-NUCA under two classes of rule sequences -- periodic and aperiodic. It is indeed revealed that characterisation of t-NUCAs with aperiodic rule sequences is more difficult than that of periodic rule sequence. In this work, we have defined a special class rule sequences, named reversible rule sequences. The periodic rule sequences are reversible rule sequences. However, it is not known till now that whether there exists any aperiodic reversible rule sequences. A future research can also be conducted to identify subclasses of aperiodic rule sequences and to characterise the t-NUCAs with the rule sequence of those subclasses.

The paper has studied surjectivity of t-NUCAs under various conditions, such as, whether $G_f$ and $G_g$ are surjective or not, whether lattice is finite or infinite, etc. It is shown that surjectivity of t-NUCAs is a decidable problem when periodic rule sequence. An algorithm has been reported to do so. However, the question is still open for aperiodic rule sequence. In case of injectivity of t-NUCAs, on the other hand, the problem is straightforward: when $G_f$ and $G_g$ are injective then the t-NUCA is injective, and vice versa.

Unlike classical CAs, bijectivity of a t-NUCA is not sufficient to show reversible behavior. Rather, a reversible rule sequence is required. The paper has characterised reversible t-NUCAs and shown that inverse of a t-NUCA is also a t-NUCA. Finally, the cyclic behavior of finite t-NUCAs have been discussed. In the transition diagram, unlike classical CAs with finite size, a configuration may appear in more than one chain of configurations. Hence, a configuration can appear in different cycles of various length. We have introduced two nomenclature for these configurations: recurrent if a configuration appears infinitely in a cycle, transient otherwise. Once again it is possible to identify recurrent and transient configurations of t-NUCA with periodic rule sequence. But, the question remains open for aperiodic rule sequences. The work on cyclic behavior of finite t-NUCAs opens up new challenges of (1) finding of the cycle structure of finite t-NUCA, and (2) synthesis of t-NUCAs with large cycles.

The t-NUCAs can be utilized in numerous applications, which can be future research directions of this work. In particular, the reversible t-NUCAs with large cycles are the promising tools in generating pseudo-random numbers.

\bibliographystyle{tcs} 
\bibliography{main}

\end{document}